# Second-harmonic optical circular dichroism of plasmonic chiral helicoid-III nanoparticles

Florian Spreyer[†,1], Jungho Mun[†,2], Hyeohn Kim[†,3], Ryeong Myeong Kim[3], Ki Tae Nam[*,3], Junsuk Rho[*,2,4], and Thomas Zentgraf[*,1]

[1]Department of Physics, Paderborn University, Warburger Straße 100, 33098 Paderborn, Germany

[2]Department Mechanical Engineering, Pohang University of Science and Technology (POSTECH), Pohang 37673, Republic of Korea

[3]Department of Materials Science and Engineering, Seoul National University, Seoul 08826, Republic of Korea

[4]Department of Chemical Engineering, Pohang University of Science and Technology (POSTECH), Pohang 37673, Republic of Korea

[*]Corresponding author

[†]These authors contributed equally to this work.

## ABSTRACT

While plasmonic particles can provide optical resonances in a wide spectral range from the lower visible up to the near-infrared, often symmetry effects are utilized to obtain particular optical responses. By breaking certain spatial symmetries, chiral structures arise and provide robust chiroptical responses to these plasmonic resonances. Here, we 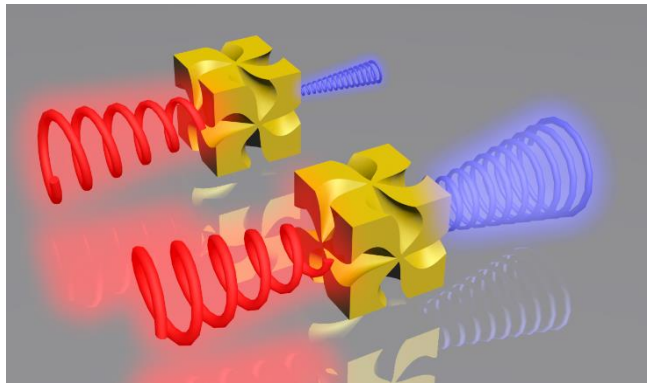 observe strong chiroptical responses in the linear and nonlinear optical regime for chiral L-handed helicoid-III nanoparticles and quantify them by means of an asymmetric factor, the so-called g-factor. We calculate the linear-optical g-factors for two distinct chiroptical resonances to −0.12 and −0.43 and the nonlinear optical g-factors to −1.45 and −1.63. The results demonstrate that the chirality of the helicoid-III nanoparticles is strongly enhanced in the nonlinear regime.

During the last decade, plasmonic nanoparticles (NPs) have become of great interest in widespread research fields, especially in health, material science, and optics. [1-4] The progress of different nanofabrication methods, *e.g.* high-resolution lithography, molecular self-assembly, and amino-acid- and peptide-direct synthesis nowadays enable more complex designs of plasmonic NPs. [5-10] The reams of options in different sizes, shapes, materials, and arrangements open up a variety of applications, such as protein detectors, solar cells, optoelectronics, or surface-enhanced Raman spectroscopy. [11-15] However, each application requires nanoparticles with unique optical properties. Especially plasmonic NPs often need well-tailored properties to utilize the localized surface plasmons resonances (LSPR's), which many optical applications rely on.

The optical properties of plasmonic NPs are highly dependent on the size and shape of their surfaces. [16-19] While the volume-to-surface ratio of NPs is comparatively low, effects that are localized at the surface become more important. Hence, for tailored optical applications, the precise control of these properties is as essential as well as the characterization of them. The geometrical and optical properties of tailored NPs are often investigated by imaging technologies such as scanning electron microscopy or optical measurements, *e.g.* spectrometry, transmission-/absorption measurements, or more recently scatterometry. [4, 13, 20] Depending on the particular application, some properties might be more important than others. For example, due to the low volume-to-surface ratio of NPs, the absorption of plasmonic NPs is less important than the scattering properties when it comes to optical chirality.

Chiroptical properties, as such as reported in this article, often rely heavily on LSPR's. Although many of the mentioned methods are widely used for the characterization of the linear optical properties of chiral plasmonic nanostructures, [7, 21-23] plasmonic NPs can also offer pronounced nonlinear optical properties arising from LSPR's. The strong nonlinear response of plasmonic nanostructures is of great interest for applications in active optical systems like emitters as well as sensors [24-29] and various studies investigated a variety of properties and applications. [30-32]

Here, we provide a study of the linear and nonlinear optical responses in individual L-handed chiral plasmonic gold NPs. A schematic illustration of the investigated L-handed helicoid-III NP is shown in Figure 1A. Based on our experiments, we calculate the linear optical g-factor from measured scattering spectra to −0.43. Furthermore, we study the impact of this chiroptical response on the second harmonic generated (SHG) signal. We found that the corresponding nonlinear g-factor is increased to −1.63, which is almost 4 times stronger than compared to the linear chiroptical response.

## RESULTS AND DISCUSSION

**Nanoparticle structure**

A schematic illustration of the investigated L-handed helicoid III NP is shown in Figure 1A. Each face of the cube is implemented with four chiral arms evolving away from the center point on each side with increasing thickness resulting in curvature gaps. The NP's have a broken mirror and inversion symmetry and belong to the 432-point group. After the synthesis, the NP's are spin-coated on a quartz substrate where the centrifugal forces separate the NP's spatially. The separation prevents near-field interactions, which can have a significant impact on the linear and nonlinear chiroptical response of the plasmonic NP's. [33]

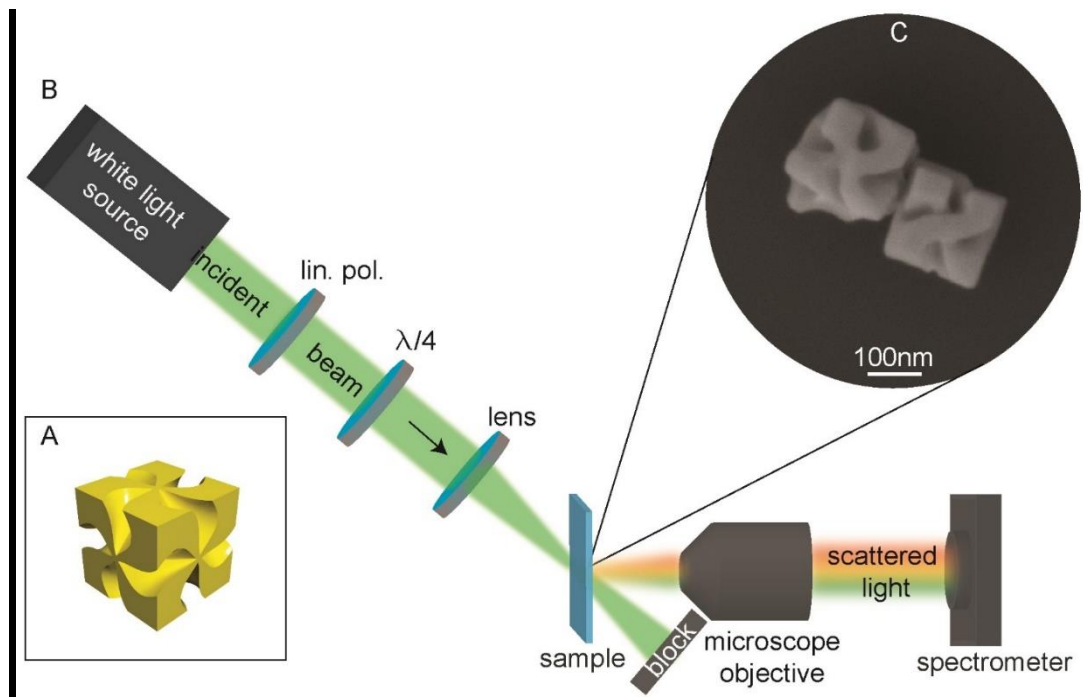

**Figure 1: Measuring the chiroptical response of chiral nanoparticles.** (A) Schematic illustration of the investigated NP with L-handed chirality. (B) Schematic illustration of the measurement setup for obtaining the linear scattering spectra. The directly transmitted beam is not collected by the microscope objective and blocked after passing the sample such that only the scattered light is measured either at a spectrometer or camera. (C) Scanning electron microscopy (SEM) image showing two chiral NPs.

Figure 1C shows a scanning electron microscopy image of the grown NP. As expected, the NPs show the desired chiral structure on its surface where four chiral arms evolve from the center of the side. Although many NPs obtain the aspired surface chirality, minute structural deviations may appear arising from a non-uniform growth. Such artifacts can be of various kinds, *e.g.* size, shape, and underdeveloped arms which will result in divergent chiroptical responses. [20] However, the superior chirality of these NPs might be still retained for most structural deviations, which is of high importance for the linear and nonlinear optical regime. For further terminology, the concept of chirality describes the average NP with a clear (but not necessarily defect-free) chiral structure.

**Linear-optical circular dichroism**

For analyzing the NP's chiroptical response and its strength, we first calculate the power of the absorbed and scattered light ($P_A$ and $P_S$) with full-wave Maxwell simulations using a commercial finite element method solver (COMSOL Multiphysics) in a wavelength range of 500-1600 nm. Therefore, we consider a chiral gold NP with edge lengths of L=150-190 nm in a homogeneous host medium with a refractive index of $n_h = 1.3$, which is illuminated by left- or right-circular polarized plane wave incident fields. The optical properties of gold are taken from Johnson and Christy.[34] $P_A$ and $P_S$ are obtained by integrating the local field energy dissipation of total fields and the Poynting vectors of scattered fields, respectively (see Figure 2A-B). Due to the relatively large size of NP, absorption is comparatively low for all wavelengths, while scattering is more present for near-infrared wavelengths, starting from around 800 nm. This behavior is observable for either, LCP or RCP polarization, and differences between the total extinction spectra of both polarizations are barely visible. To visualize any differences in the extinction spectra ($P_E = P_A + P_S$) for the different polarization states, we quantize the chiroptical response of the chiral NP, introducing the so-called linear asymmetry g-factor $g_T = 2 \cdot \frac{P_{LCP}^E - P_{RCP}^E}{P_{LCP}^E + P_{RCP}^E}$, where $P_{LCP/RCP}^E$ represent the calculated extinct power for LCP or RCP input polarization. Figure 2C illustrates the calculated g-factors $g_T$ for different NP edge lengths, showing, that for different particle sizes the local minimum in g-factor moves to greater wavelengths, indicating that slightly different NP geometries can have a great impact.

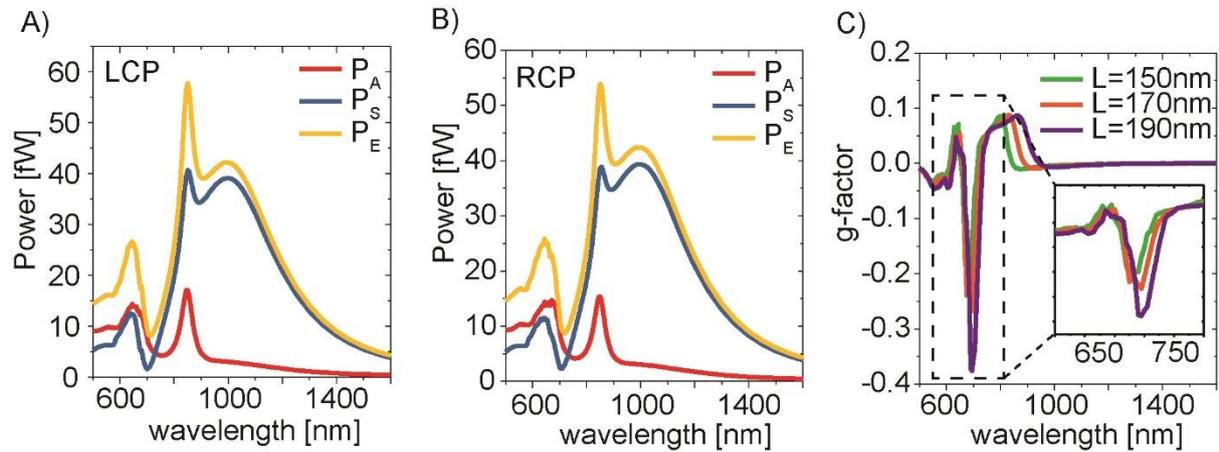

**Figure 2:** Full-wave numerical **simulation results for the chiral NP.** (A) Calculated absorption-, scattering- and extinction spectra ($P_A, P_S$ and $P_E$) for LCP input polarization. (B) Calculated absorption-, scattering- and extinction spectra ($P_A, P_S$ and $P_E$) for RCP input polarization. (C) Linear g-factor $g_T$ calculated for different NP edge lengths via extinction spectra $P_E$ with LCP/RCP polarization, shown in (A) and (B). The inset highlights the local minimum in g-factor at around 700 nm, marked by the dashed box.

The linear-optical g-factor, obtained by our numerical simulations, shows a clear chiroptical response at around 700 nm, where a g-factor of almost −0.4 implies, that a higher amount of RCP intensity is scattered/absorbed by the NP (Figure 2C). Further, for near-infrared wavelengths 1100–1600 nm, both LCP and RCP light are scattered equally, so that the g-factor remains at the value near zero, implying no chiroptical response is present in this spectral region. Therefore, we focus in the following on measuring the chiroptical response in the visible wavelength range.

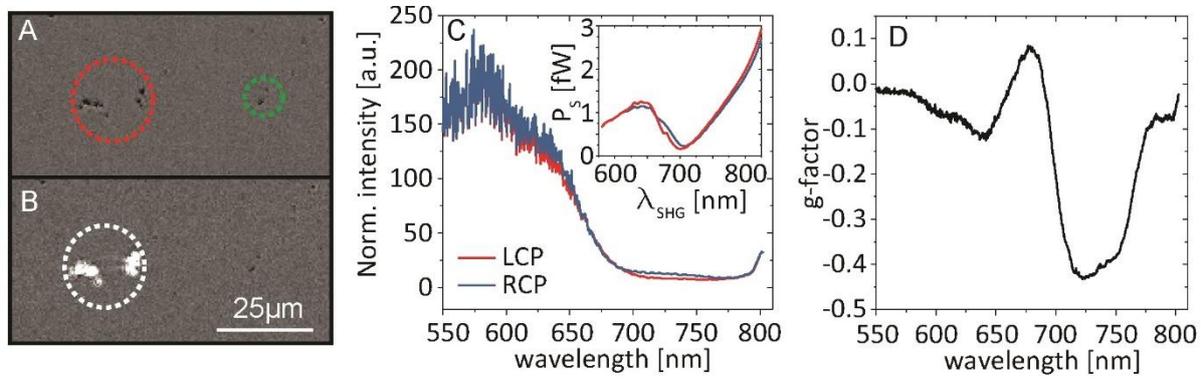

**Figure 3: Linear optical g-factor measurement.** (A) Optical microscopy image of several NPs gathered in colloids (red) and single or few NP (green). (B) Scattering image of the same sample area as in panel A for illumination by the white-light laser source. The illumination area is indicated by the white dotted circle. The light that is scattered at the chiral NPs appears as bright spots. (C) Obtained scattering spectra for LCP/RCP illumination. Note that these spectra do not correspond to the scattered light shown in (B). The scattering spectra are normalized on the source spectrum. The inset shows as a comparison the theoretical scattering spectra from Figure 2A-B for the different circular polarizations. (D) Calculated linear g-factor obtained from the measured scattering spectra in panel (C).

The linear chiroptical response is measured by a scattering setup, which is shown in Figure 1B. The sample is illuminated with the light of a supercontinuum white-light laser source (Fianium WL-400) with wavelengths of 600–800 nm. After passing a linear polarizer, the polarization state of the illuminating beam is converted to LCP or RCP by using a broadband quarter-wave plate, as the investigation of chiroptical responses requires circularly polarized light. For clarification, we use the definition of LCP and RCP polarization state from the source point of view with the propagating beam. Subsequently, a lens focuses down the circularly polarized beam with a few milliwatts of average laser power onto the NP sitting on a glass substrate. The incident angle of the illumination beam is chosen to be greater than the numerical aperture of the microscope objective (40×/NA 0.6) that collects the scattered light of the NPs. In this way, only scattered light is measured by the spectrometer and the transmitted beam is blocked behind the sample, resulting in a greatly suppressed background signal. The spectrometer consists of a grating monochromator (Andor Kymera 193i) with a low noise charged-coupled device camera (Andor iDus DU420A-BVF). A microscopy image of the sample surface with the chiral NPs, obtained with the same setup but using a CMOS camera for imaging instead of the spectrometer, is shown in Figure 3A. Due to the spin coating process, it is covered with colloids of NPs as well as single or few NPs. By illuminating the NPs with the laser source, light from the NPs is strongly scattered (depending on the individual geometry and symmetry) and measured on our detection system. An example image of the scattered light is shown in Figure 3B, where the illumination area is chosen such that only a small area of NPs is illuminated. Note that the scattered light in Figure 3B results from a colloid of NPs for better visibility in the image and is not used for the subsequently determined g-factors since we want to investigate NP with less near-field interaction than possible. Figure 3C shows typical scattering spectra for a single NP. Note that the scattered intensity is normalized to the illumination spectrum of the light source. From the measured linear scattering spectra, a difference in the scattered intensity for each input polarization is already noticeable. By comparing the experimental results of the scattered intensity to the theoretically obtained results shown in the inset, we find that the simulations resemble the experimental results, although some differences are observable. The main observation is, that both scattering spectra show a comparatively low scattering efficiency for wavelengths around 700 nm. Although the local minimum in the experimentally obtained spectra shows a slight redshift, this can be explained by a small difference in particle size compared to the simulation. In addition, the dip is broadened, which can result from the different preconditions of the simulation and experiment and inhomogeneity's of the particle shape itself. The simulated scattering spectra are based on the overall scattering cross-section, whereas in the experiment, the NP is excited only for one angle and the microscope objective collects only the light scattered into a certain solid angle. In addition, possible defects in the NP structure also play a role in its scattering behavior, which are not considered in the simulation. Nevertheless, the agreement of the simulation with the experimental results becomes more visible, when the g-factor is calculated.

To investigate and assess the chiroptical response of the presented NP, we normalize the chiroptical response similar to the theoretical calculations and express it again as the convenient asymmetry g-factor, given by $g_L = 2 \cdot \frac{I^\omega_{LCP}-I^\omega_{RCP}}{I^\omega_{LCP}+I^\omega_{RCP}}$, wherein $I^\omega_{LCP/RCP}$ represent the intensities of the scattered light by LCP/RCP input polarization. In Figure 3D, we plot the g-factor, which was calculated by the scattering spectra shown in Figure 3C. The result shows two local minima for the g-factor at wavelengths of 640 nm and 723 nm. Compared to previous studies, the local minimum at 723 nm with a g-factor value of −0.43 provides clear evidence of a linear chiroptical response arising from the chirality of the plasmonic NP. The negative value of the g-factor means that significantly more RCP light is scattered than for the LCP state. The strong difference in the scattering at this wavelength results from the excitation of the LSPR for RCP light. The second local minimum at 640 nm shows an additional but weaker chiroptical response with a g-factor value of $g_L$ = −0.12. As the simulation (see Figure 2C) shows only one chiroptical response in the visible wavelength range for L-handed helicoid-III NP is expected, previous studies have shown that structural anomalies can have a great impact on the chiroptical response of individual NP [20]. Measuring the chiroptical response of a NP with a structural anomaly can differ from the expected chiroptical response for NP with perfect geometries. Such anomalies can be some kind of structural defect, *e.g.* deformation, or NPs gathered in a small colloid of at least two NPs. NPs with structural anomalies exhibit complex chiroptical responses and therefore, additional chiroptical features due to the coupling of additional complex LSPR modes can appear. But still, as mentioned before, the presence of artifacts does not invalidate the concept of chirality and the loss of chiroptical responses. By comparing the g-factors of the simulations and the experiments, a good agreement can be observed, where a strong chiroptical response for wavelengths around 700-750 nm is present. Surprisingly, at these wavelengths only a weak scattering of the NP is observable, both in the theoretically and experimentally obtained results. Nevertheless, the measurements and simulations show the same strength in chiroptical response, which is quantified by the g-factor of about $g_L = -0.4$.

**Nonlinear-optical circular dichroism**

Breaking certain symmetries opens up a variety of nonlinear optical effects related to the excitation of LSPR's in plasmonic nanostructures. It is well-known, that the properties of the nanostructures can significantly impact the nonlinear response. [35-36] In the case of the presented chiral NP's the mirror and inversion symmetry are broken and a strong nonlinear signal arising from the NP is expected. [32] For measuring this nonlinear response, we extended the setup from Figure 1B for additional illumination with a short pulse laser. The extended transmission setup for the nonlinear measurement is schematically depicted in Figure 4A. Hereby, the nonlinear response in the regime of 600-800 nm is of particular importance, since the LSPR`s provide a linear chiroptical response. To utilize these resonances for the nonlinear answer, it is necessary to excite them in the near-infrared regime at 1200–1600 nm. Although theoretical calculations, shown in Figure 2C, do not show a chiroptical response for near-infrared wavelengths, we expect a chiral response for the SHG of NIR wavelengths due to the plasmonic chiral resonances at wavelengths of 600–800 nm. To measure the SHG signal of the NPs in the wavelength range of 600–800 nm, we use an optical parametric oscillator (OPO) as a tunable near-infrared light source. While the spectral width of the coherent infrared light source is only approximately 20 nm, we vary the illumination wavelength in steps of 10 nm from 1200–1600 nm and measure the SHG spectrum for each illumination wavelength and polarization state. After the laser beam passes the polarization optics generating the circular polarization states (LCP or RCP), the non-infrared wavelengths get filtered out by a long pass before reaching the sample. Note that the laser beam is focused on the same sample spot as the previously used white-light laser source for obtaining the scattering spectra. In this way, we ensure to measure the response of the same NP, whose linear chiroptical response is measured in the first place. To visualize the measurement, Figure 4B shows a microscopy image of the SHG signal arising from the same NPs that are highlighted in Figure 3B by a white circle. As expected, the plasmonic NPs emit a strong SHG signal. To investigate a nonlinear chiroptical response to compare it to the linear results, the SHG spectra for each illumination wavelength in the near-infrared ranging from 1200-1600 nm for LCP and RCP input polarization are measured for the same NP as before.

Figure 5C shows selected SHG spectra, measured for 1220 nm and 1390 nm pump wavelength and LCP/RCP input polarization. As already visible, the spectra, measured for the same wavelength but opposite helicity, do not show the same signal strength, which already gives a hint on a nonlinear chiroptical answer similar to the linear chiroptical answer shown in Fig. 3D. Evaluating the nonlinear optical response for better comparison to the linear g-factor, we define $g_{NL} = 2 \cdot \frac{I^{2\omega}_{LCP} - I^{2\omega}_{RCP}}{I^{2\omega}_{LCP} + I^{2\omega}_{RCP}}$, as the nonlinear asymmetry g-factor, in analogy to the linear asymmetry g-factor $g_L$, wherein $I^{2\omega}_{LCP/RCP}$ represent the SHG intensities for the LCP/RCP input beam polarization. The results of the nonlinear chiroptical response, represented by $g_{NL}$ calculated for the SHG signals, obtained for each pump wavelength are shown in Figure 5D. Note that the determined nonlinear g-factor is not related to the SHG signal shown in Figure 4B, but measured for a single NP and not the cluster the SHG signal is arising from.

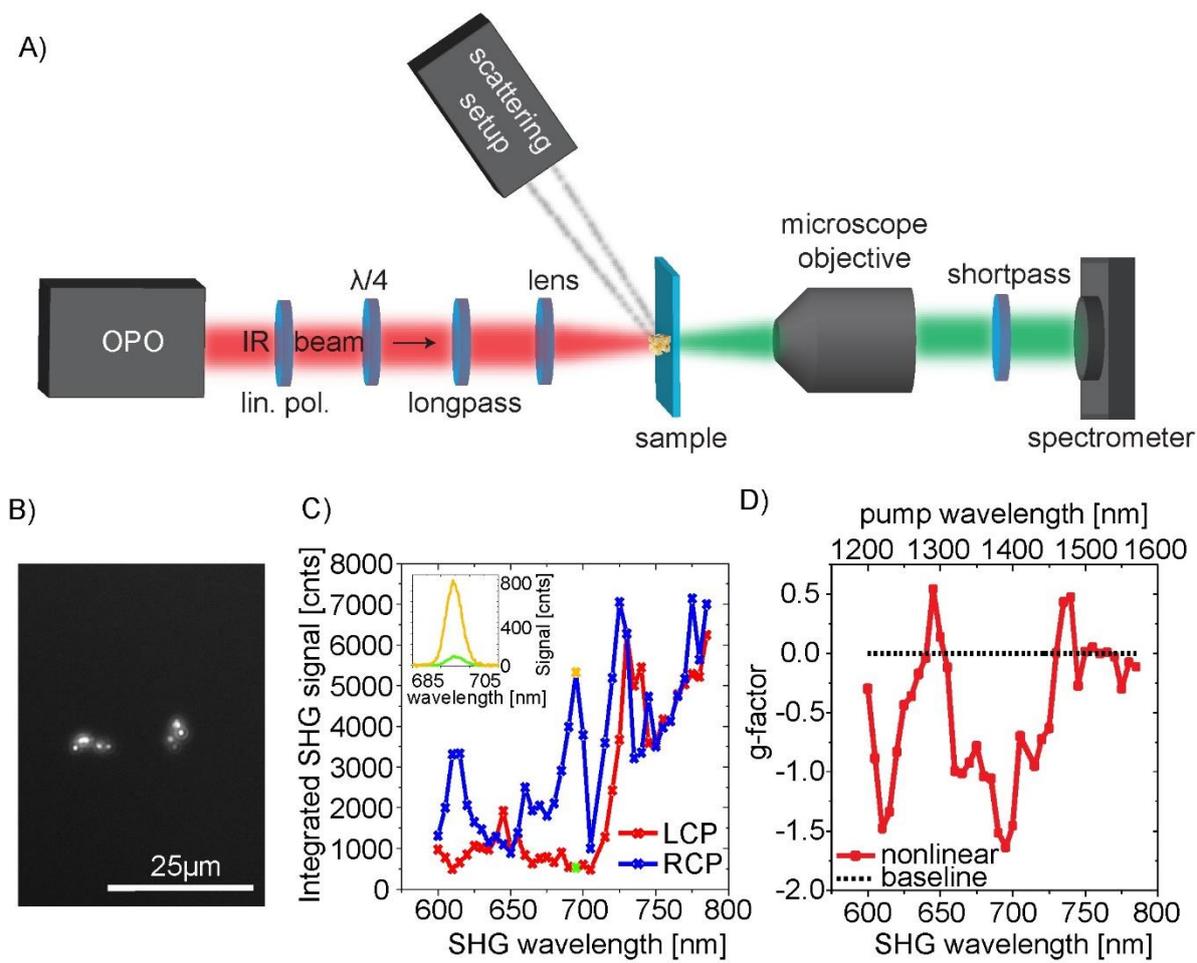

**Figure 4: Results for the nonlinear g-factor.** (A) A coherent laser beam, generated by an OPO, initially passes the polarization optics for generating the circular polarization input and a subsequent long pass to filter any non-infrared light. A lens focuses the beam onto the NP and the generated SHG is collected by a microscope objective. The SHG is separated from the NIR wavelengths by a short pass filter and measured by a spectrometer. (B) Microscopy image of SHG generated by NP shown in Figure 3B (red circled). (C) Measured SHG signals for varying illuminating wavelengths with either LCP or RCP polarization. The inset shows the SHG spectra for LCP and RCP polarization at illumination wavelengths of 1390 nm, marked in green and yellow. (D) Calculated nonlinear g-factor for the measured nonlinear SHG spectra with LCP/RCP input. A baseline is added as a reference for NP without any chirality that would result in a g-factor of zero.

By evaluating the nonlinear chiroptical response with the help of $g_{NL}$, two main differences between the linear and the nonlinear g-factor stand out. First, although we still see two local minima in the nonlinear g-factor, they are not precisely located at the same wavelength that we observed for the linear chiroptical response. While the linear g-factor has its local minima at around 640 nm and 723 nm, we find the local minima for the nonlinear chiroptical response at around 610 nm and 695 nm, slightly blue-shifted by around 30 nm. Although this effect is not fully understood yet, it might arise due to the comparatively high extinction of the illuminating wavelengths in the NIR range. A previous study has shown, that by increasing the extinction at illumination wavelengths, shorter than twice the intended second harmonic wavelength can result in a blueshifted second harmonic signal, although plasmonic modes predict the emission at longer wavelengths.[37] The second important difference between the linear and the nonlinear g-factors concerns the strength of the nonlinear chiroptical response. Compared to $g_L$, the values for $g_{NL}$ reach even lower values, indicating a stronger nonlinear chiroptical response. While the linear g-factor at the wavelength of 723 nm is given by −0.43, the corresponding nonlinear g-factor for the local minimum at 695 nm reaches −1.63, which is almost a factor of 4 larger. While this already states a significant impact of chirality on the nonlinear generation, the second local minimum of the nonlinear g-factor (located at 610 nm) reaches a value of −1.45, which results in an even higher ratio of more than 12. This effect can be explained by the strong localized near-fields, induced by the incident fundamental light, generating SHG light and coupling to the chiral surface plasmon modes. While the SHG intensity is square correlated to the E-field strength of the illuminating NIR light, this nonlinear correlation can therefore lead to a distinct chiroptical response, since the difference between the SHG signal of each polarization is increased. In addition, the strong localized near-fields in plasmonic nanostructures are highly dependent on its shape and a qualitative description of the increased nonlinear chiroptical response in complex structures is non-trivial and would need knowledge of the exact shape of the NP including every anomaly. The formation of localized field hot spots at such anomalies combined with the nonlinear correlation of incident and SHG E-fields can result in an even greater difference of SHG signal between the polarization states, leading to an increased chiroptical response. A theoretical approach, where the nonlinear chiroptical response is modeled in a simulation, as well as discussion about different local modes providing the chiroptical response in such chiral NPs can be found in the supplementary material.

As such anomalies can be different for each nanoparticle, we measure the linear and nonlinear g-factors of further particles to support our observations. The results are shown in Figure 5 for three more NPs. From the linear g-factors of the three NPs, shown in Figure 5A-C (top), we find that all g-factors show small differences compared to each other and the linear g-factor of the previous NP (NP0, shown in Figure 3D). Although each of them provides a strong circular dichroism, where more RCP than LCP light is scattered, resulting in a g-factor lower than zero, the local minima in the g-factors are not located at the same wavelength and exhibit a different value. This can be explained by the slightly different shape and size in each NP. Since every NP has its individual form due to the imperfections of the fabrication process, each one provides a different scattering response, which is highly dependent on size and defects. This observation is supported by the theoretical results, shown in Figure 2C, illustrating the impact of the NP size on the linear g-factor as well as on recently published data of different linear g-factors for various NPs. [20]

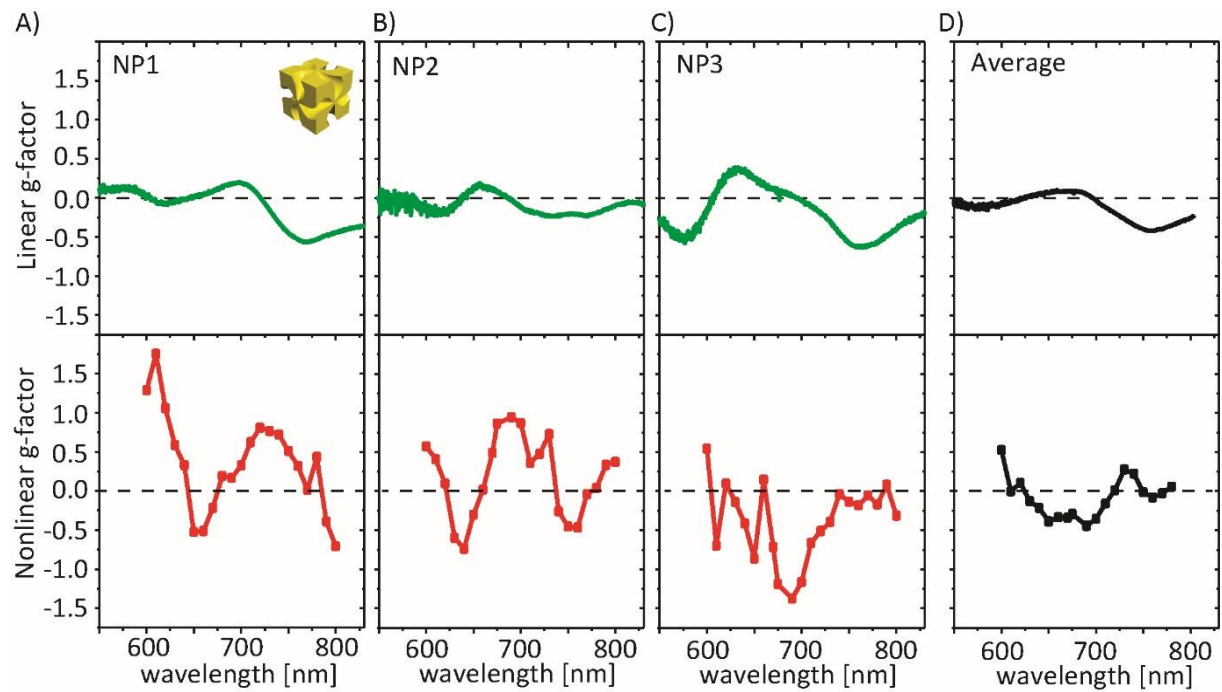

**Figure 5: Linear and nonlinear g-factors of different NPs.** (A-C) Linear (top) and nonlinear (bottom) g-factors for NP1-3. The inset in (A) shows a schematic illustration of the L-handed helicoid-III NP. (D) Average linear (top) and nonlinear (bottom) g-factor for the NP shown in A-C, Fig. 3D and Fig. 4D. A baseline (dotted line) is added as a reference for NPs without any chirality that would result in a g-factor of zero.

The corresponding nonlinear g-factors for each NP are shown in Figure 5A-C (bottom panel) and differ from each other like they do in the linear case. Nevertheless, the increase in g-factor for the nonlinear case can still be observed for each of the three NPs and is determined to the factors of 6.6, 3.4 and 2.3 for the NP1-3, respectively. As the local minima in the nonlinear g-factor of the NP0 is blue-shifted, this shift is not visible for the NP1-2, but rather a small red-shift is present. One explanation can be an incorrect relative assignment of the local minima from the linear g-factor to the local minima of the nonlinear g-factor, where a further local minimum in the linear g-factor is present for wavelengths greater than 800 nm due to defects. Also, an increased absorption of the NP at the fundamental wavelength for LCP light can lead to a spectral shift of the circular dichroism. This mechanism has been reported previously and describes a spectral shift for SHG intensities away from the plasmon resonances if the absorption at the fundamental wavelength is increased.[37] As the linear and nonlinear response of our NPs is highly dependent on their individual shape, distortions and defects can lead to an increased absorption at the fundamental wave influencing the circular dichroism in the nonlinear case of SHG. Nevertheless, the increase in chiroptical response for all NPs is still given. If one assumes a red-shift for the local minima, an increase of the chiroptical response of about 3 can be measured. On the other hand, a considered blue shift results in an increase of only 1.1. As each NP provides chiroptical responses at different wavelengths, an ensemble measurement can weaken or cancel the dichroism. This means, that the relative increase in the chiroptical response needs to be considered for each particle individually. Overall, we conclude from our observations that the increased chiroptical response can be observed for individual NPs, which arises from the nonlinear dependence of the SHG signal from the E-field strength of the illuminating light. Further, the individual shape of each NP, as well as defects, plays an important role when it comes to nonlinear chiroptical responses.

As the measurements of the linear and nonlinear g-factors of L-handed helicoid-III NPs have shown an increase in their chiroptical response, we repeat the measurement for NPs with the opposite helicity. The experimentally obtained linear and nonlinear g-factors for three D-handed helicoid-III NP's are shown in Figure 6, whereas a schematic illustration of the D-handed helicoid-III NP is shown in Figure 6A.

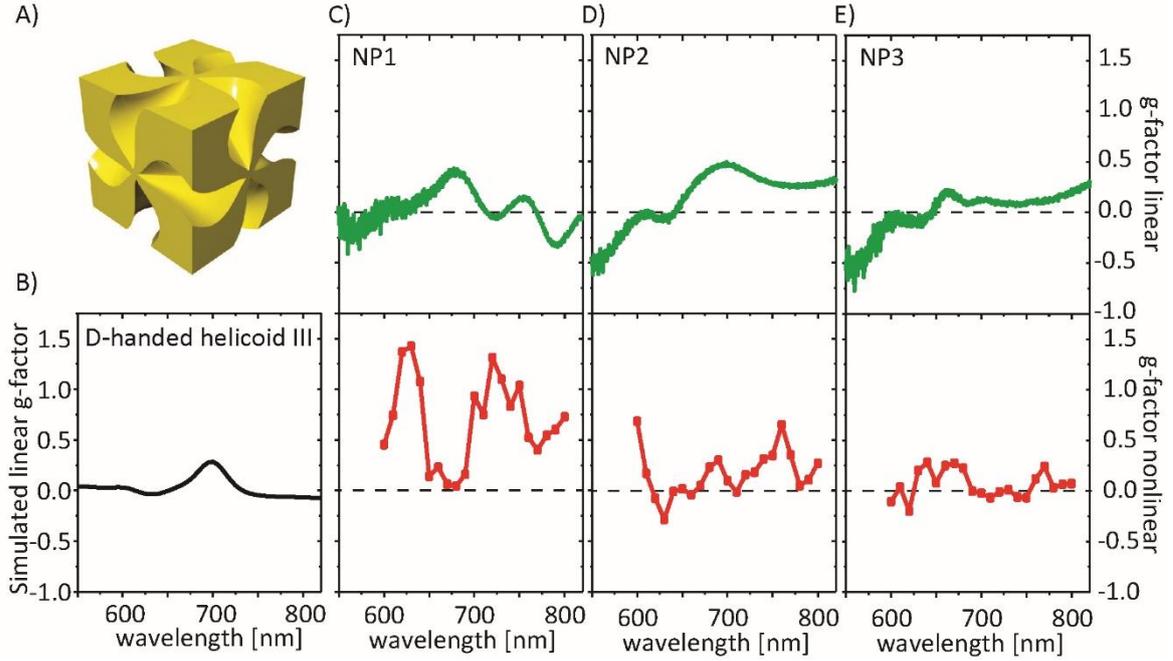

**Figure 6: Linear and nonlinear g-factors of D-handed helicoid-III NPs.** (A-C) Linear (top) and nonlinear (bottom) g-factors for NP1-3. The inset in (A) shows a schematic illustration of the D-handed helicoid-III NP. (D) Average linear (top) and nonlinear (bottom) g-factor for the NP shown in A-C. A baseline (dotted line) is added as a reference for NP without any chirality that would result in a g-factor of zero.

For the NPs with the opposite helicity, a similar behavior for the different circular polarization states is expected: as for the L-handed helicoid-III NP, more RCP light is scattered, the D-handed helicoid-III NP should scatter more LCP light. Hence, it is expected that the linear g-factor for D-handed helicoid-III NPs, calculated by $g_L = 2 \cdot \frac{I^\omega_{LCP} - I^\omega_{RCP}}{I^\omega_{LCP} + I^\omega_{RCP}}$, is mirrored at the baseline compared to the g-factor of the L-handed helicoid-III NP. This is supported by the calculated g-factor based on simulation results shown in Figure 6B, obtained in the same way as the simulation results in Figure 2C. We found, that at wavelengths, where previously local minima were observed for L-handed helicoid-III NPs, now local maxima for D-handed helicoid-III NPs are observed.

By looking at the experimental results of the D-handed helicoid-III NP, the determined linear g-factors (Figure 6C-E, top panels) also show strong circular dichroism with different distributions and local maxima at different wavelength positions. This observation matches the results obtained for the L-handed helicoid-III NPs, shown in Figure 5. The g-factor has predominantly positive values, which is expected for this helicity, where more RCP than LCP light is scattered. Further, the curves peak at wavelengths slightly lower than 700 nm, hinting, that the measured D-helicoid-III NPs are slightly smaller than the L-helicoid-III NPs of opposite helicity. As the simulation illustrates, it is expected, that these linear g-factors only show one predominant local maximum. Clearly, this is not the case for each NP where additional maxima might arise due to imperfections in the growth process, leading to a distorted nanoparticle. Nevertheless, the nonlinear g-factors still can be determined, which are shown in Figure 6C-E (bottom panel). First, the results show a similar behavior compared to the linear g-factors: the nonlinear g-factors have predominantly values greater than zero. However, the previously stated increase in the g-factor is not as strong as for the NP of the other helicity and is also slightly shifted in their spectral response similar to the L-handed helicoid-III NPs. By comparing the nonlinear g-factors to the linear g-factors, it seems that for the NP1 and NP3, the local maxima of the nonlinear g-factors undergo a small blue-shift in respect to their linear counterparts, whereas for NP2 a red-shift is visible. In respect to this shift, the increase in the g-factor and therefore the chiroptical response can be quantified by the factors of 3.5, 1.3, and 1.2 for the NP1-3, respectively.

Note, that the linear g-factor for achiral seeds, which were used in a seed solution for the fabrication of the presented chiral NPs, can be found in the supplementary material. As these seeds have a present inversion symmetry, they don't show any SHG signal and therefore, the calculation of a nonlinear g-factor is not possible. Nevertheless, their achiral optical response supports our observations, made in L- and D-handed helicoid-III NPs.

## CONCLUSION

In summary, we present theoretical calculations for the linear chiroptical response over a wide wavelength range from 600-1600 nm and underline them with experimental results for the linear and nonlinear chiroptical response of several chiral plasmonic NPs. By measuring the scattering spectra, we confirm a linear chiroptical response in the wavelength range of 600-800 nm, where all particles show a strong circular dichroism, whose characteristics are based on their chirality. Further, we examined the nonlinear optical response by measuring the SHG signal of the same chiral NPs, by illuminating it with wavelengths from 1200-1600 nm. Although our simulations do not show any chiroptical response for the near-infrared wavelengths ranging from 1000–1600 nm, a strong nonlinear chiroptical response in the SHG signal is measured. Similar to the linear chiroptical response, we observe distinct minima in the nonlinear chiroptical response with slightly spectrally shifted positions. In addition, we found that the strength of the nonlinear optical chirality is even larger than for the linear case with up to an increase by a factor of 12. However, this increase differs for every nanoparticle and we conclude, that it is highly dependent on the individual shape of each nanoparticle including any distortion or defect, which impacts its chiroptical response. Nevertheless, we link the strong nonlinear chiroptical response to the linear counterpart, which is the visible range and originates from the excitation of localized surface plasmon polaritons. We presented data of the linear and nonlinear for nanoparticles for both L- and D-handed helicoid-III NPs and show, that the linear and nonlinear chiroptical response switches depending on their chirality. Although the D-handed helicoid-III particles show a weaker increase in chiroptical response, it might result from further difficulties in the fabrication process. Our study demonstrates that the chirality of the NPs becomes more pronounced by the nonlinear frequency conversion, which might be a suitable tool for analyzing weaker chiralities.

## METHODS

The NPs are grown in a L- or D-glutathione and a seed solution, wherein the asymmetric growth due to the chiral glutathione molecule leads to chiral morphology evolution. Octahedral seeds were synthesized as previously reported, [15] centrifuged (6.708 $g$, 150 s) twice, and dispersed in 1 mM hexadecyltrimethylammonium bromide (CTAB). The growth solution for chiral nanoparticles was prepared by adding 0.8 mL of 100 mM CTAB and 0.2 mL of 10 mM gold chloride trihydrate into 3.95 mL of deionized water to form a $[AuBr_4]^-$ complex. $Au^{3+}$ was then reduced to $Au^+$ by the rapid injection of 0.475 mL of 100 mM ascorbic acid. The growth of chiral nanoparticles was started with the addition of 5 μL of 5 mM L-glutathione solution and 50 μL of seed solution into the growth solution. The temperature was maintained at 30°C in a water bath for 2 h. The particle solution was centrifuged twice (1.677 $g$, 60 s) to be dispersed in a 1 mM CTAB solution for further characterization.

## ACKNOWLEDGEMENTS

The authors acknowledge the funding provided by the European Research Council (ERC) under the European Union's Horizon 2020 research and innovation program (grant agreement No. 724306), the German Academic Exchange Service (DAAD, Project ID 57572468) and GEnKO program (NRF-2021K2A9A2A15000174) funded by the National Research Foundation of Korea (NRF). J.R. acknowledges the NRF grants (NRF-2019R1A2C3003129, CAMM-2019M3A6B3030637, NRF-2019R1A5A8080290) funded by the Ministry of Science and ICT (MSIT) of the Korean government. K.T.N. acknowledges the support from Creative Materials Discovery program through the NRF funded by the MSIT (NRF-2017M3D1A1039377), and the LGD-SNU Incubation program funded by LG Display. J.M. acknowledges POSTECH PIURI fellowship, and the NRF postdoctoral fellowship (NRF-2021R1A6A3A01087429) funded by the Ministry of Education of the Korean government.